# Resting State ASL : Toward an optimal sequence duration

Corentin Vallée[1], Pierre Maurel[1], Isabelle Corouge[1], and Christian Barillot[1]

[1]*Univ Rennes, Inria, CNRS, Inserm, IRISA UMR 6074, VISAGES ERL U-1228, F-35000, Rennes, France*

## Synopsis

**Resting-state functional Arterial Spin Labeling (rs-fASL) in clinical daily practice and academic research stay discreet compared to resting-state BOLD. However, by giving direct access to cerebral blood flow maps, rs-fASL leads to significant clinical subject scaled application as CBF can be considered as a biomarker in common neuropathology. Our work here focuses on the link between overall quality of rs-fASL and duration of acquisition. To this end, we consider subject self-Default Mode Network (DMN), and assess DMN quality depletion compared to a gold standard DMN depending on the duration of acquisition.**

## Introduction

In both resting-state functional MRI (rs-fMRI) academic research and clinical applications, Arterial Spin Labeling (ASL) remains rare compared to Blood Oxygenation Level Dependent (BOLD) fMRI. Despite ASL lower estimated signal-to-noise ratio compared to BOLD-fMRI, rs-fASL is a promising valuable technique to be used in clinics since :

- It provides cerebral blood flow (CBF), and thus leads to potential significant clinical applications as CBF can be used as a biomarker in various neuropathology (Alzeihmer disease, severe depression...).
- For over the last decade ASL processing has rapidly improved, and becomes more and more disseminated for clinical usage.

As asked in[1] : "What ASL time series length would be adequate?", our purpose is to provide a first answer to this question. While some indications were given for rs-BOLD[2], this work has not been done yet for rs-fASL, as far as we know. In this abstract, we will focus on one functional brain network characterizing the resting-state: the Default Mode Network (DMN). Given a duration of acquisition (DOA), we make the assumption that the quality of the DMN defines quite well the quality of the corresponding rs-fASL acquisition. As we want to work on a subject-scaled (first-level) analysis for further applications and as a preliminary work, we will describe the variation of quality of the subjects self-DMN compared to a gold standard reference.

## Methods

We considered a dataset of 6 healthy subjects who volunteered for a 24min40s (426 volumes) rs-fASL acquisition. We split each acquisition into 20 parts of various durations (5min to 24min by 1). Each of the 6x20 acquisitions were processed independently with typical preprocessing[1]. In order to work at subject-scaled level (i.e. considering each subject DMN, not group built DMN), we relied on Seed-Based Analysis (SBA). We selected seeds in the medial prefrontal cortex as they give the best results, according to the number of non-noised DMN exhibited among the 120 acquisitions. We built maps of the second order approximation of the FDCR estimator of correlation $\hat{r}_{adj.} = \hat{r}\left(1 + \frac{1-\hat{r}}{2n}\right)$ as we can get rid of the bias of the canonical estimator of correlation $\hat{r}$ considering the given number of $n$ volumes (86 for 5min up to 412 for 24min). Then, we tested whether each voxels signal is positively correlated with the seed (risk FWER-corrected set at 0.01). Let's the binary mask obtained be called $bmap_{doa,s}$ with the subject $s$. We wanted to compute the Jaccard index between the 120 bmaps with a gold standard reference, let's call it $J_s(bmap_{doa,s}, ref_s)$. However, taking a subject's DMN as gold standard leads to an overwhelming dependency between bmaps and the reference. Moreover, we can not build a global $J(bmap_{doa,s}, ref_s)$ this way as $ref_s$ would depends on $s$. Hence, we took as gold standard the DMN from the recent atlas of functional area of the brain (MSDL) proposed by Varoquaux et al.[3]. Now, for every subject $s$, we can compute $J_s(bmap_{doa}, ref)$ and as MDSL DMN is not subject dependent, we can also compute a global $J(bmaps_{doa}, ref)$. Finally, we modelled the Jaccard index for each subject and for all subjects with the non-parametric local regression LOESS[4] with two degrees local polynomials and a span equal to 0.8.

## Results

Every $J_s$ and its associated LOESS is shown in Figure 1. Some corresponding overlays of bmaps and MSDL DMN on MNI152 template are shown in Figure 2 to provide qualitative illustrations of Jaccard Index. The global $J$ and its associated LOESS is presented in Figure 3.

1. All $J_s$ increase before around 16-18min.
2. After a peak around 17 min each $J_s$ except $J_1$ and $J_6$ decrease.
3. For $J_1$, $J_3$, $J_5$, close DOA does not imply close Jaccard Index.
4. Global $J$ increases till 17min.
5. After 17min $J$ decrease.
6. Considering the LOESS on global $J$, on average, $J$ has 95% chance to be higher at 17 minutes than before 12 minutes and after 21 minutes.

## Discussion

Surprisingly, the quality of the DMN looks to be not maximal when DOA is maximal. In our dataset, the optimal quality of rs-fASL seems to be reached around 17 minutes. Result (3.) suggests quality may not be stable over time in subject-scaled analysis.

## Conclusion

In this preliminary work, we exhibit a subject-scaled and an overall tendency about quality of rs-fASL, depending on DOA (under the assumption of correlation between quality of the DMN and overall quality of the rs-fASL acquisition). For subject-scaled analysis, checking subset of a whole acquisition should be considered as quality seems not to be always stable. In further work, we would like to confirm the tendency by evaluating other usual functional areas of interest like motor, visual or language.

## Acknowledgements

MRI data acquisition was supported by the Neurinfo MRI research facility from the University of Rennes I. Neurinfo is granted by the the European Union (FEDER), the French State, the Brittany Council, Rennes Metropole, Inria, Inserm and the University Hospital of Rennes.

## Figures

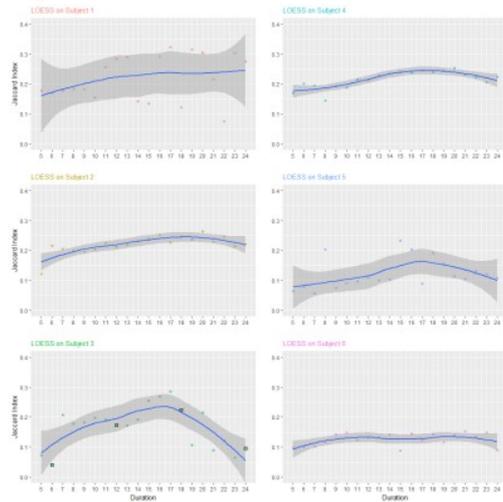

Figure 1 : Curves of the computed LOESS for each subject. The grey area is the 95% confidence area related to the LOESS. Framed dots are corresponding to overlays shown in Figure 2. Colour on subjects as well as aspect ratio are matching those in Figure 3.

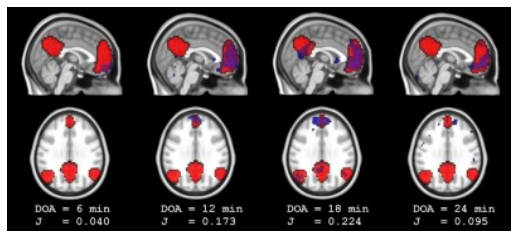

Figure 2 : Some illustrations of values of the Jaccard Index according to the DOA with corresponding maps. All images shown are registered in the MNI 152 template. Red areas correspond to the MSDL DMN. Blue areas correspond to subject voxels whose signal were significantly positively correlated to the chosen seed signal in medial prefrontal cortex.

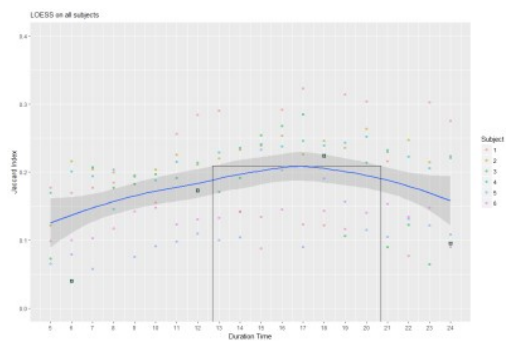

Figure 3 : Curve of the computed LOESS for all subject. The grey area is the 95% confidence area related to the LOESS. Framed dots are corresponding to overlays shown in Figure 2. Colour on subjects as well as aspect ratio are matching those in Figure 1. Black lines show the average values of Jaccard index according to the LOESS at 95% chance to be higher than before 12 minutes and after 21 minutes.